\documentclass{article}%
\usepackage{amsfonts}
\usepackage{amsmath}
\usepackage{amssymb}
\usepackage{graphicx}%
\begin{document}

\title{Embedding G\"{o}del's universe in five dimensions }
\author{J. B. Fonseca-Neto$^{a}$, C. Romero$^{a}$ and F. Dahia$^{b}$\\$^{a}$Departamento de F\'{\i}sica, Universidade Federal da Para\'{\i}ba,\\C.Postal 5008, 58051-970 Jo\~{a}o Pessoa, Pb, Brazil\\E-mail: jfonseca@fisica.ufpb.br; cromero@fisica.ufpb.br\\$^{b}$Departamento de F\'{\i}sica, Universidade Federal de Campina\\Grande, 58109-970, Campina Grande, Pb, Brazil\\E-mail: fdahia@df.ufcg.edu.br }
\maketitle

\begin{abstract}
According to the Campbell-Magaard theorem, any analytical spacetime can be
locally and analytically embedded into a five-dimensional pseudo-Riemannian
Ricci-flat manifold. We find explicitly this embedding for G\"{o}del's
universe. The embedding space is Ricci-flat and has a non-Lorentzian signature
of type $(++---)$. We also show that the embedding found is global.

\end{abstract}

\section{Introduction}

In 1949, Kurt G\"{o}del \cite{Godel} found a solution of Einstein's field
equations which soon became very popular because it described a spacetime
possessing very strange properties. Among these was the existence, predicted
by the model, of timelike closed curves which violate global causality. Even
though it is not viable as a physical model of our universe, G\"{o}del's
solution has some historical importance as it certainly stimulated a great
deal of research on questions of causality and global properties of
relativistic spacetimes \cite{Roger,Geroch}.

Due to its peculiarity, different aspects of the so-called G\"{o}del's
universe have always been studied with interest. For example, Rosen
\cite{Rosen} in 1965, was able to characterize G\"{o}del's model as a
four-dimensional hypersurface embedded in a pseudo-Euclidean space with ten
dimensions. An interesting result concerning the embedding of the G\"{o}del
spacetime in higher dimensions has been published which shows that the
G\"{o}del metric may also be viewed as a squashed anti-de Sitter geometry,
thereby allowing the construction of a global algebraic isometric embedding in
seven-dimensional flat spaces \cite{Rooman}. Motivated by the current interest
in higher dimensional theories of gravity, particularly the five-dimensional
Randall-Sundrum models\ with non-compact extra dimension \cite{Randall}, the
embedding of G\"{o}del and G\"{o}del type solutions on a 3-brane has been
studied recently \cite{Modgil}.

Further motivation to study the embedding of G\"{o}del%
\'{}%
s solution comes from the so-called induced-matter proposal \cite{wesson}.
According to this proprosal any solution of Einstein's equations may be
obtained from five-dimensional vacuum field equations, with matter in
four-dimensions being generated or "induced" by purely geometrical means.
Following this scheme, Wesson and colaborators have shown how to obtain from
five-dimensional vacuum (or Ricci-flat) spaces a number of known solutions of
the Einstein equations (regarded as hypersurfaces in five dimensions) whose
energy-momentum tensor is generated by the extra-dimension \cite{Wesson}. In
fact, the energy-momentum thus generated corresponds to the extrinsic
curvature of the four-manifold embedded in five-dimensional vacuum space
\cite{Maia}. It has been later realised \cite{Romero-Tavakol} \ that any
energy-momentum tensor can be generated in this way, provided that any
solution of Einstein%
\'{}%
s equations has an embedding into a five-dimensional Ricci-flat solution, and
this is almost precisely the content of an embedding theorem of differential
geometry\cite{Campbell,Magaard}. Therefore, according to the this theorem
(proposed by Campbell (1926) and later proved by Magaard (1963)), it is
possible, to locally embed G\"{o}del's solution in a five-dimensional
Ricci-flat pseudo-Riemannian space \cite{Campbell,Magaard}. In the light of
the induced-matter theory, that means it must be possible to geometrically
generate a source of matter and energy which is the source of G\"{o}del's
universe with all its peculiarities. Let us note that conjectural
considerations of such embeddings in the context of the induced-matter theory
were already put forward some years ago \cite{Reboucas}.

Let us recall then the content of the Campbell-Magaard theorem
\cite{Campbell,Magaard}. It states that any $n$-dimensional
\ pseudo-Riemannian manifold $(M^{n},g)$ can be locally, analytically and
isometrically embedded in a Ricci-flat (n+1)-dimensional manifold
$(N^{n+1},\widetilde{g)}$. Since its "rediscovery" in the nineties
\cite{Romero-Tavakol}\ the theorem has found \ a number of applications and
has been discussed in various contexts in the literature
\cite{Rippl,Lidsey,Agnese,Aderson,Dahia1,Dahia2,Seahra,Dahia3,Dahia4,Chervon,Anderson2,Dahia5}%
. Therefore, in view of the Campbell-Magaard theorem one would like to look at
G\"{o}del's solution as a hypersurface embedded in a five-dimensional
Ricci-flat space.

\section{The embedding of G\"{o}del's universe in ten dimensions}

In a classical paper, in which he performs the embedding of various
relativistic spacetimes, Rosen exhibits an embedding of G\"{o}del's universe
in a ten-dimensional pseudo-Euclidean flat space $E^{10}$\ with signature
$(+++++-----)$ by explicitly giving the embedding transformations. Let us see
how it is done.

G\"{o}del's metric may be expressed in the form \cite{Godel}%
\begin{equation}
ds^{2}=a^{2}(dt^{\ast}+e^{x^{\ast}}dy^{\ast})^{2}-dx^{\ast2}-\frac{1}%
{2}e^{2x^{\ast}}dy^{\ast2}-dz^{\ast2} \label{Godel1}%
\end{equation}
where $a$ is a constant. Let us choose a coordinate system $\{Z^{1}%
,...,Z^{10}\}$\ of $E^{10}$\bigskip, in which the line element is given by
\begin{align}
dS^{2}  &  =(dZ^{1})^{2}+(dZ^{2})^{2}+(dZ^{3})^{2}+(dZ^{4})^{2}+(dZ^{5}%
)^{2}-\nonumber\\
&  -(dZ^{6})^{2}-(dZ^{7})^{2}-(dZ^{8})^{2}-(dZ^{9})^{2}-(dZ^{10})^{2}%
\end{align}

\bigskip Now let us consider the four-dimensional hypersurface $\Sigma_{4}%
$\ defined by the parametric equations
\[
Z^{1}=at^{\ast}%
\]%
\[
Z^{2}=(a/\sqrt{2})e^{x^{\ast}}\cos y^{\ast}%
\]%
\[
Z^{3}=(a/\sqrt{2})e^{x^{\ast}}\sin y^{\ast}%
\]%
\[
Z^{4}=a\sqrt{2}e^{\frac{1}{2}x^{\ast}}\cos\frac{1}{2}(t^{\ast}+y^{\ast})
\]%
\[
Z^{5}=a\sqrt{2}e^{\frac{1}{2}x^{\ast}}\sin\frac{1}{2}(t^{\ast}+y^{\ast})
\]%
\[
Z^{6}=ax^{\ast}%
\]%
\[
Z^{7}=az^{\ast}%
\]%
\[
Z^{8}=(a/\sqrt{2})e^{x^{\ast}}%
\]%
\[
Z^{9}=a\sqrt{2}e^{\frac{1}{2}x^{\ast}}\cos\frac{1}{2}(t^{\ast}-y^{\ast})
\]%
\[
Z^{10}=a\sqrt{2}e^{\frac{1}{2}x^{\ast}}\sin\frac{1}{2}(t^{\ast}-y^{\ast})
\]
A simple calculation shows that the metric induced in $\Sigma_{4}$ by the
pseudo-Euclidean metric of $E^{10}$ is equal to G\"{o}del's metric
(\ref{Godel1}).

The above result may be viewed as an application of the well-known
Janet-Cartan theorem \cite{Janet,Cartan}\ which claims that any $n$%
-dimensional\ Riemannian space $M^{n}$\ can be locally and isometrically
embedded in an Euclidean space $E^{m}$\ with $m\leq n(n+1)/2$ dimensions$.$ In
fact, as G\"{o}del's metric is pseudo-Riemannian, one should consider the
extension of Janet-Cartan theorem to pseudo-Euclidean spaces, obtained by
Friedman \cite{Friedman}. In this extension the following condition is
required for the embedding to take place: if $ds^{2}=g_{\alpha\beta}%
dx^{\alpha}dx^{\beta}$ and $dS^{2}=\eta_{ab}dZ^{a}dZ^{b}$ denote the line
elements of $M^{n}(p,q)$ e $E^{m}(r,s)$, respectively, with $p$ and
$q$\ $(n=p+q)$ being the number of positive and negative eigenvalues of
$g_{\alpha\beta}$, \ $r$ and $s$ $(m=r+s)$ the number of positive and negative
eigenvalues of $\eta_{a_{b}},$ then $r\geq p$ and $s\geq$ $q$ . Thus, if
$n=4$, the maximum number of dimensions required for the embedding space is
$m=10$, which means that for performing the embedding of G\"{o}del's spacetime
Rosen needed this maximum number .

Before we proceed, let us note that, by defining the new coordinates
$t=at^{\ast},x=ax^{\ast},y=ay^{\ast},z=az^{\ast}$ , with $a=1/\sqrt{2w},$ the
line element (\ref{Godel1}) can be put in the form \cite{Hawking-Ellis}
\begin{equation}
ds^{2}=dt^{2}-dx^{2}+\frac{1}{2}\exp(2\sqrt{2}wx)dy^{2}-dz^{2}+2\exp(\sqrt
{2}wx)dtdy \label{Godel2}%
\end{equation}

\section{The embedding of G\"{o}del's universe in five dimensions}

In this section we shall show how to obtain the embedding of G\"{o}del's
spacetime in a five-dimensional Ricci-flat space, the metric of which has
signature $(+---+)$. We already know that when \ $n\geq3$,\ \ the
Campbell-Magaard alows us to lower the number of dimensions of the embedding
space $N^{n+1}$ from $n(n+1)/2$ to $n+1$, as long as $N^{n+1}$ be Ricci-flat.
However, we shall not employ directly the Campbell-Magaard; instead we shall
make use of the following theorem due to Magaard \cite{Magaard}:

\textbf{Theorem (Magaard). } Let $(M^{n},g)$ be a $n$-dimensional
pseudo-Riemannian manifold, $\{x^{\mu}\}$ a local coordinate system of a
neighbourhood $U$ of $p\in M^{n}$, with coordinates $(x_{p}^{1},...,x_{p}%
^{n})$ defined by the parametrization $\mathbf{x:}U\rightarrow M^{n}$. A
sufficient and necessary condition for $(M^{n},g)$, with line element
$ds^{2}=g_{\alpha\beta}(x)dx^{\alpha}dx^{\beta}$, to be locally, isometrically
and analytically embedded in a $(n+1)$-dimensional manifold $(M^{n+1}%
,\widetilde{g})$ is that there exist analytical functions
\begin{equation}
\overline{g}_{\alpha\beta}=\overline{g}_{\alpha\beta}(x^{1},...,x^{n},x^{n+1})
\label{métricateorema1}%
\end{equation}%
\begin{equation}
\overline{\phi}=\overline{\phi}(x^{1},...,x^{n},x^{n+1}) \label{fi}%
\end{equation}
definided in an open set $D\subset\mathbf{x}(U)\times\mathbb{R}^{n}$
containing the point $(x_{p}^{1},...,x_{p}^{n},0)$, satisfying the following conditions:%

\[
\overline{g}_{\alpha\beta}(x^{1},...,x^{n},0)=g_{\alpha\beta}(x^{1}%
,...,x^{n})
\]
in an open set of $\mathbf{x}(U)$; $\overline{g}_{\alpha\beta}=\overline
{g}_{\beta\alpha}$, $\left\vert \overline{g}_{\alpha\beta}\right\vert \neq0 $
; $\overline{\phi}$ $\neq0$, and that
\begin{equation}
d\overline{s}^{2}=\overline{g}_{\alpha\beta}dx^{\alpha}dx^{\beta}%
+\varepsilon\overline{\phi}^{2}dx^{n+1}dx^{n+1} \label{coordenadasmagaard}%
\end{equation}
with $\varepsilon^{2}=1$, represents the line element of $M^{n+1}$ in a
coordinate neighbourhood $V$ of $M^{n+1}$\cite{Magaard,Dahia}.

In the light of the above theorem let us take $n=4$, $\varepsilon=1$,
$\phi=-k^{2}$, where $k$ is a constant, and the set of analytical functions
$\{\overline{g}_{\alpha\beta}(t,x,y,z,\psi\}$ \footnote{We can choose the
range of the new coordinates to be given by $-\infty<t,x,y,z,\psi<-\infty$.},
$(\alpha,\beta=0,1,2,3)$ the non-null elements of which are $\overline{g}%
_{00}=1;\overline{g}_{02}=\overline{g}_{20}=\exp(\sqrt{2}w(x+k\psi
));\overline{g}_{11}=-1;\overline{g}_{22}=\frac{1}{2}\exp(2\sqrt{2}%
w(x+k\psi));\overline{g}_{13}=\overline{g}_{31};\overline{g}_{33}=-1$. Clearly
the conditions $\overline{g}_{\alpha\beta}=\overline{g}_{\beta\alpha}$,
$\overline{\phi}$ $\neq0$ are satisfied, and also $\left\vert \bar{g}%
_{\alpha\beta}\right\vert =-\frac{1}{2}\exp(2\sqrt{2}w(x+k\psi))\neq0$.
\ Moreover, $\overline{g}_{\alpha\beta}(t,x,y,z,0)=g_{\alpha\beta}(t,x,y,z)$,
hence the functions $g_{\alpha\beta}$ may be identified with the components of
G\"{o}del's metric written in the form (\ref{Godel2}). We conclude, therefore,
from the above theorem that the G\"{o}del's universe can be embedded in a
five-dimensional space $M^{5}$ with metric given by
\begin{equation}
dS^{2}=dt^{2}-dx^{2}+\frac{1}{2}\exp(2\sqrt{2}w(x+k\psi))dy^{2}-dz^{2}%
+2\exp(\sqrt{2}w(x+k\psi))dtdy+k^{2}d\psi^{2} \label{Godel3}%
\end{equation}
the embedding taking place for $\psi=0$, i.e. by choosing the embedding
functions given by $t\rightarrow t,x\rightarrow x,y\rightarrow y,z\rightarrow
z,\psi=0.$

If we calculate \footnote{This can be done quickly and efficiently by
employing algebraic computation programs such as SHEEP or GRTensor.} the
components $^{(5)}R_{ab}$\ of the Ricci tensor directly from (\ref{Godel3}) we
get $^{(5)}R_{ab}$\ $=0$. We see then that the five-dimensional manifold
$M^{5}$, in which G\"{o}del's universe appears as the hypersurface $\psi=0,$
is a Ricci-flat space, and that proves our claim.

Let us conclude this section by noting two points. The first is that the
manifold on which G\"{o}del's metric (\ref{Godel2}) is defined is $R^{4}$,
i.e. $-\infty<t,x,y,z<\infty$ \cite{Hawking-Ellis},\ and it is clear that the
present embedding takes the whole of $R^{4}$ into $M^{5}$, irrespective of the
domain chosen for $\psi$. Moreover, we see that the embedding functions and
the metric of the embedded spacetime are analytic in $R^{4}$ while the metric
of embedding space is analytical in $M^{5}.$\ It turns out then that in spite
of the local character of the theorem mentioned previously in this particular
case the embedding found happens to be \textit{global }( a global version of
the Campbell-Magaard theorem\ has been discussed recently in
\textit{\cite{Nikolaos})}. \ It is interesting to have a look at the
components of the extrinsic curvature tensor $\Omega_{\alpha\beta}$ of the
hypersurface $\ \psi=const$\ of $M^{5}.$ In the coordinates of (\ref{Godel3})
it can easily be shown that $\Omega_{\alpha\beta}$ is given by $\Omega
_{\alpha\beta}=-\frac{1}{2k}\frac{\partial\overline{g}_{\alpha\beta}}%
{\partial\psi}$\cite{Dahia2}, so that the nonvanishing components of
$\Omega_{\alpha\beta}$\ are $\Omega_{02}=\Omega_{20}=$\ $-\frac{1}{2}\sqrt
{2}w\exp(\sqrt{2}w(x+k\psi)),$\ $\Omega_{22}=-\sqrt{2}w\exp(2\sqrt{2}%
w(x+k\psi)).$As we see, the extrinsic curvature is also well-behaved
(analytical) everywhere for any hypersurface of the foliation $\ \psi=const,$
in particular for $\psi=0$.\ As a consequence of the global caracter of the
embedding, all global properties so characteristics of G\"{o}del%
\'{}%
s universe, such as the existence of closed timelike curves, are preserved in
$M^{5}$.

The second point is concerned with the existence of timelike curves which may
appear to be closed in the four-dimensional spacetime, hence appearing to
violate global causality, but are not closed when viewed from the perspective
of $M^{5}.$ Let us explain what we mean. In the context of the induced-matter
theory, for example, a geodesic motion in five-dimensions may appear
accelerated when viewed from five dimensions. This \textit{anomalous}
acceleration manifest itself when geodesics lying on $M^{5}$ are projected
(essentially by hiding the extra dimension) \ onto the hypersurface which
represents ordinary four-dimensional spacetime \cite{Mashhoon, Monte}. As the
projected curve need not be a geodesic with respect to the metric of the
hypersurface, there appears acceleration. Thus in this approach the projection
of a non-closed curve in $M^{5}$ may appear to be closed (when we drop the
fifth dimension) in four-dimensional spacetime, much in the same way as in
ordinary tridimensional space \ $R^{3}$ \ we can project an helix into a
circle. This fact perhaps suggests that concept of causality when extra
dimensions of the spacetime are present needs further discussion.

\section{Final remarks}

We would like to call attention for the fact that the space $(M^{5}%
,\widetilde{g})$, which is a solution of the Einstein vacuum field equations
in five dimensions, has the peculiarity of possessing a non-Lorentzian
(ultra-hyperbolic)\ metric, with two timelike dimensions. Spaces of these kind
have been studied recently, mainly in connection with the idea that massless
particles in five dimensions may appear "massive" when viewed from
four-dimensional spacetime \cite{Wessontwotimes,Youm,Youm1}. There are also
claims that two times theories may find some motivation in M-theory
\cite{Vongehr}. One must recognise, however, that from the point of view of
physics,\ ultra-hyperbolic spaces may pose some problems with respect to
causality. On the other hand, examples of embedding spaces with extra timelike
dimensions are many \cite{Rosen}, \ and include, for instance, the embedding
of the Schwarzschild spacetime in a six-dimensional flat manifold obtained for
the first time by Kasner \cite{Kasner, Kocinski}.\ Isometric embeddings in
flat spaces with two times have also been investigated in the context of
branes \cite{Herdeiro}. Finally, it is interesting to note that it is not
possible to globally embed a spacetime which is not globally hyperbolic into a
pseudo-euclidean space with only one timelike dimension \cite{Penrose}. We do
not know whether a similar result holds in the case of Ricci-flat embedding spaces.

\section{Acknowledgement}

The authors would like to thank CNPq-FAPESQ (PRONEX) for financial support.
CR\ thanks Dr. Carlos Herdeiro for helpful discussions. Thanks also go to the
referee for pointing out serious typos in the original manuscript and for
invaluable comments.

\end{document}